\documentclass[11pt]{article}
\usepackage{epsfig}
\usepackage[usenames]{color}
\usepackage{graphicx}
\usepackage{amsmath}
\def\la{\langle}\def\ra{\rangle}
\def\be{\begin{eqnarray}}\def\ee{\end{eqnarray}}
\def\lsim{\mathrel{\rlap{\lower3pt\hbox{\hskip1pt$\sim$}}
     \raise1pt\hbox{$<$}}} %less than or approx. symbol
\def\gsim{\mathrel{\rlap{\lower3pt\hbox{\hskip1pt$\sim$}}
     \raise1pt\hbox{$>$}}} %greater than or approx. symbol
\def\le{ \begin{array}{ll}}\def\re{\end{array}}

\def\lear{ \left( \begin{array}{cc}}\def\rear{\end{array} \right)}

\def\le{ \left( \begin{array}{cc}}\def\re{\end{array} \right)}

\def\bi{\bibitem}

\renewcommand{\thefootnote}{\fnsymbol{footnote}}
\begin{document}
\hfill \vbox{\hbox{}}
\begin{center}{\Large\bf Topology Change and Tensor Forces \\ for the EoS of Dense Baryonic Matter}\\[0.8cm]

{ Hyun Kyu Lee\footnote{\sf e-mail: hyunkyu@hanyang.ac.kr}
}\\
{\em Department of Physics, Hanyang University, Seoul 133-791, Korea}

{ Mannque Rho\footnote{\sf e-mail: mannque.rho@cea.fr }
}\\
{\em Institut de Physique Th\'eorique, CEA Saclay, 91191 Gif-sur-Yvette c\'edex, France \&
\\Department of Physics, Hanyang University, Seoul 133-791, Korea}

\end{center}
\vspace{0.2cm}
\centerline{\today}
\vspace{0.2cm}
\begin{center}
{\Large\bf Abstract}
\end{center}

   When skyrmions representing nucleons are put on crystal lattice and compressed to simulate high density, there is a transition above the normal nuclear matter density ($n_0$) from a matter consisting of skyrmions with  integer baryon charge to a state of half-skyrmions with half-integer baryon charge. We exploit this observation in an effective field theory framework to access dense baryonic system. We find that the topology change involved in the transition implies changeover from a Fermi liquid structure to a non-Fermi liquid with the  chiral condensate in the nucleon ``melted off." The $\sim 80\%$ of the nucleon mass that remains ``unmelted," invariant under chiral transformation, points to the possible origin of the (bulk of) proton mass that is not encoded in the standard mechanism of spontaneously broken chiral symmetry. The topology change engenders a drastic modification of the nuclear tensor forces, thereby nontrivially affecting the EoS,  in particular, the symmetry energy, for compact star matter. It brings in stiffening of the EoS needed to accommodate a neutron star of $\sim 2$ solar mass. The strong effect on the EoS in general and in the tensor force structure in particular will also have impact on processes that could be measured at RIB-type accelerators.

%\pacs{ }

%\maketitle
\vfill

\pagebreak
\setcounter{footnote}{0}
\renewcommand{\thefootnote}{\arabic{footnote}}
%%%%%%%%%%%%%%%%%%%%%%%%%%%%%
\section*{Dedication}
Long before effective field theory anchored on chiral perturbation began to play a predominant role in nuclear physics -- and with a large success, Gerry Brown and one of the authors (MR) started to ask in what way chiral symmetry figured in nuclear physics, in particular (Gerry) in nuclear forces and (MR) in exchange currents. This part of the story is recounted in the Gerry Brown Festschrift volume~\cite{festschrift}.  One of the early observations among many that we made then was that it could figure particularly importantly in the structure of nuclear tensor forces~\cite{tensorforce}. And this is what led us to the proposal of what is now referred to as ``Brown-Rho scaling" (or ``BR scaling" for short). In 23 years that have elapsed since then, a surprising, totally unanticipated, twist in the structure of the tensor forces was discovered, which, if confirmed to be correct, promises to have a novel and profound implication on dense nuclear matter, particularly for the EoS for compact stars. Gerry did not participate in this new development but we are  certain that what is described in this note would have pleased him immensely.

We dedicate this note --prepared for a contribution to ``EPJA Special Volume on Nuclear Symmetry Energy" -- to Gerry Brown.

\section{Introduction}
%%%%%%%%%%%%%%%%%%%%%%%%%%%%%
In constructing EoS for compact stars, it is commonly assumed in going beyond  normal nuclear matter density, $n_0=0.16$ fm$^{-3}$, that one is dealing with nucleonic matter in the Fermi liquid state. This is the assumption that underlies a variety of nuclear models employed in the literature generically anchored on density functionals, among which the popular Skyrme potential model~\cite{skyrmepot}. Both Walecka-type relativistic mean field theory (RMFT)~\cite{walecka} involving, in addition to nucleons (or baryons in general), vector mesons and scalar mesons~\cite{matsui} {\it and} chiral Lagrangians containing four-fermi field operators taken at mean field~\cite{gelmini-ritzi}, belong to the same class of models. They turn out to be fairly successful at near nuclear matter density because they are equivalent to Landau Fermi liquid fixed point theory~\cite{shankar,polchinski} with the parameters of the effective quasiparticle Lagrangian ``marginal" with vanishing beta functions in the ``large N" limit~\footnote{Here $N\propto k_F$, where $k_F$ is the Fermi momentum.}. Away from the fixed point but in its vicinity, one can then endow the parameters with smooth density dependence and {\it limit} to two-body interactions. This can describe fairly well nuclear matter around $n_0$ possessing thermodynamic consistency~\cite{songetal}.

However it is not at all obvious how to go far beyond $n_0$ as required if one wants to describe what happens at a density relevant to the interior of compact stars, such as neutron stars, quark stars, hybrid stars etc.

One currently popular approach is to write an effective Lagrangian with appropriate symmetries in terms of what are considered to be the relevant degrees of freedom in the given density regime, i.e, pions, vector mesons, scalar mesons, and baryons (nucleons and hyperons), including multi-dimension field operators and then to apply the mean field approximation. This approach has been surprisingly successful for finite nuclei as well as for infinite nuclear matter~\cite{reviewRMF}. In accessing higher densities, this mean-field approximation is simply extended without justification.

Applying the mean-field approximation to densities higher than that of nuclear matter might  be justified provided the Fermi-liquid picture continued to hold up to the density one drives the system to. However, there is no reason to expect that it will not break down as density increases: It could in fact do so if there were phase changes at increasing densities. For instance, new degrees of freedom such as condensation kaons or equivalently hyperons could emerge at higher density~\cite{kaon-hyperon}. This could involve certain order parameters signally the change of  the symmetries involved.  But there can also be changes of matter that cannot be characterized by local order parameters, a possibility that has currently attracted a great deal of attention in condensed matter physics but thus far not in nuclear/hadron physics.

In this paper, we address the state of matter that arises due to the change of topology as density increases above the normal nuclear matter density $n_0$. We suggest that when baryons are considered as solitons, that is, skyrmions, an ubiquitous concept~\cite{multifacet},  there can be a phase change that does not belong to the standard Ginzburg-Landau-Wilson paradigm, with no identifiable local order parameters, and that could signal changeover from a Fermi liquid structure to a non-Fermi liquid structure triggered by a topology change. The question we will address is whether topology can provide a qualitatively different information, not evident in topology-less formulations, that can be implemented in an effective field theory framework and be used for quantitative calculations. The procedure we take is then to ``translate" what is given in the skyrmion formulation {\it with} topology -- on crystal -- to the structure of an effective Lagrangian {\it without} topology, with which one can do systematic calculations. The assumption we make here is that topology change can be ``translated" to change in the parameters of the Lagrangian, in a spirit perhaps analogous to what's being discussed in the context of quantum mechanics~\cite{wilczek}.
\section{Hidden Local Symmetry}
%\subsection{The Lagrangian}
We consider baryons described as solitons. The soliton we are dealing with, skyrmion, is a topological object in a theory with meson fields only. The Lagrangian that gives rise to the skymion should be an effective one that is as close as possible to QCD with its chiral symmetry manifested appropriately in the energy regime we are interested in. The Lagrangian commonly taken up to date, such as the Skyrme Lagrangian, is a highly truncated one, anchored, for instance, on the large number-of-color ($N_c$) limit and other approximations, the validity of which is poorly justified from first principles. Even limited to pion fields only, there can be an infinite number of derivative terms valid in the large $N_c$ limit. Even worse, there is no reason why vector mesons, at least the lowest-lying ones $\rho$ and $\omega$,  not to mention the infinite tower, are ignored (or even integrated out)  if the vector meson masses can be counted at the same chiral order as the pion mass as is the case when one is considering matter at high temperature and/or at high density~\cite{HY:PR}. But the major stumbling block is the proliferation of the number of uncontrollable parameters in the effective Lagrangian. For instance, when only the lowest-lying vector mesons $\rho$ and $\omega$ are incorporated, limited up to ${\cal O}(p^4)$ in derivative counting -- to which the Skyrme quartic term belongs,  there are more than fourteen independent terms. It is clearly meaningless to pick a few terms -- not to mention only one, namely, the Skyrme term -- out of  so many without any guidance from theory and/or experiments.

Fortunately a recent development from holographic QCD  can improve the situation a lot better although one must admit, it is still far from realistic. Starting from the Sakai-Sugimoto model~\cite{SS} of holographic QCD in 5D which is found to give a fairly good description of the nucleon  in terms of an instanton~\cite{HRYY}, one can integrate out of the infinite tower in the Sakai-Sugimoto action the higher-lying vector mesons, leaving only the lowest vector mesons $\rho$ and $\omega$ as hidden local gauge fields~\footnote{ Whether this integrating-out procedure is correct in the bulk gravity sector is not clear.}. Keeping up to ${\cal O}(p^4)$ in the derivative expansion, one obtains an HLS Lagrangian with {\it all} parameters of the Lagrangian fixed by two physical quantities, the pion decay constant $f_\pi$ and the $\rho$-meson mass $m_\rho$~\cite{maetal1}. This miraculous simplification results thanks to a ``master formula" that gives in terms of the two constants all the coefficients of the ${\cal O}(p^4)$ Lagrangian. It is significant that the master formula holds even for the 5D YM action in curved space arrived at bottom-up with the hidden gauge fields ``emerging" from low energy chiral dynamics~\cite{son-stephanov}.

There are a few serious caveats in this procedure that we should point out. The Lagrangian so obtained is valid only in the limit that both the $N_c$ (number of colors) and the $\lambda=g_c^2 N_c$ (t' Hooft constant) tend to infinity. Neither is infinite in any sense in Nature: Numerically $N_c$ is 3 and for phenomenology, $\lambda$ is of order 10. The instanton mass or equivalently the skyrmion mass resulting from this Lagrangian goes like $\sim N_c \lambda$. If one were to work with the leading order in both $N_c$ and $\lambda$, the space would then be flat and the $\omega$ meson would be decoupled. The instanton size would then shrink to a point. In this limit, the standard collective-quantization of the soliton which gives the $1/N_c$ correction goes haywire, leading to a totally absurd splitting between the nucleon and its rotational excitation $\Delta$~\cite{maetal1}.   It is only at the next order in $\lambda$, i.e., ${\cal O}(\lambda^0)$, that the $\omega$ meson couples to give a finite size to the soliton and make the collective quantization sensible. Thus $1/\lambda$ corrections make qualitatively important effects for baryon structure. What about higher order $1/\lambda$ corrections? We do not know.

There is also the problem of $1/N_c$ corrections. The HLS Lagrangian we have, valid to the chiral order ${\cal O}(p^4)$,  is of leading order in $N_c$. There must be loop corrections that are of the same chiral order but subleading in $N_c$. Such corrections are, however, expected to be quantitatively important for certain hadron structure. For instance in the Skyrme model with pions only, the Casimir energy which contributes at subleading order, ${\cal O} (N_c^0)$, comes out to be $\sim 1/3$ of the leading order mass in magnitude. Unfortunately it is not known how to compute loop corrections in the bulk (gravity) sector in which the HLS Lagrangian we are working with is defined.\footnote{Elegant and sometimes powerful though it may be, the large $N_c$ consideration fails in providing the properties of nuclear matter. In fact, in the large $N_c$ limit, nuclear matter as it is known does not exist. There have been a flurry of papers discussing ``cold nuclear matter" in the large $N_c$ and large $\lambda$ limits based on gravity-gauge duality, none of which, as far as we know, have any resemblance to Nature. When it is pointed out that the theory does not agree with Nature, an answer offered is that ``it is Nature's problem, not the theory's."}

Given these caveats which seem to present serious obstacles, one may wonder, how can one exploit such a theory? What we are suggesting is that it is the topological structure that can be exploited independently of dynamical details.
\section{Half-Skyrmions}
The key observation that led to our reasoning is that when skyrmions are put on crystal lattice to simulate many-nucleon systems, the energetically most favored state is one in which the skyrmions fractionize into half-skyrmions~\cite{goldhaber-manton,manton-sutcliffe1}. This feature is more or less independent of the crystal symmetry involved. Our basic premise in what follows is that this can be taken as established.  Now when averaged over the single shell, the chiral condensate $\sigma\propto \la\bar{q}q\ra$ (where $q$ is the chiral quark massless in the chiral limit) is found to go to zero although it is not locally zero. This property is also generic, independently of the detailed structure, symmetry and dynamics, of the Lagrangian used for the skyrmion. This is a topology change, later identified as a sort of phase transition.  The resulting half-skyrmion matter  possesses an enhanced symmetry that arises at certain high baryon density~\cite{manton-sutcliffe1,park-vento}. Where that density is located depends on dynamical details of the Lagrangian and will constitute one of the most crucial practical issues for phenomenology considered in this paper. It is remarkable that the presence of such a topology change in the scheme we are adopting -- which could be free of dynamical complications -- is found to have a big impact on the properties of the EoS of dense matter.
\subsection{The skyrmion-half-skyrmion transition density $n_{1/2}$}
In order for such a topology change to be relevant to the structure of baryonic matter, it must not be too high in density, for if it were too high, then first of all it could not be accessed experimentally and secondly the HLS Lagrangian we are dealing with could not be trusted. It should not be too low relative to normal nuclear matter density either, for if it were too low, then it would be in conflict with experimental data that are accurately known. It should therefore be above $n_0$ but not too far above~\cite{dongetal}. There is a hint from the $A=4$ nucleus described as the $B=4$ skyrmion that the classical configuration is of 8 half-skyrmions~\cite{manton-sutcliffe}, which could imply that the classical half-skyrmion structure could already be present in the alpha particle.

To locate theoretically the transition density $n_{1/2}$ in dense matter, we take the baryon HLS Lagrangian (BHLS for short) described above~\cite{maetalcrystal}. It has the right degrees of freedom to control the location of $n_{1/2}$. Both vector mesons are found to be equally important. First of all, without them, the density $n_{1/2}$ comes much too low to be realistic. With the $\rho$ but without the $\omega$, however, it lies much too high. On the other hand,  scalar mesons, the only degrees of freedom that are not controlled by hidden local symmetry, have very little influence~\cite{park-vento}. With both the $\rho$ and $\omega$ included,  the reasonable density range comes out to be
\be
1.5\lsim n_{1/2}/n_0\lsim 2.0.
\ee
Within this range the results we obtain are robust.
\subsection{Intrinsic parameter changes at $n_{1/2}$}
We now look at what takes place to baryonic matter at $n_{1/2}$. As noted, the changeover at that density  has no apparent local order parameter. We will see that it can be associated with qualitative changes in the EoS, that will be attributed to a changeover from Fermi liquid to non-Fermi liquid. Here we shall discuss the effects on the parameters of the hadrons propagating on the skyrmion background provided by the HLS Lagrangian. Skipping the details, we summarize the results from \cite{maetalcrystal}:
\begin{enumerate}
\item The in-medium pion decay constant $f_\pi^*$ figuring in the {\it effective} HLS Lagrangian is found to drop proportionally to\footnote{The constant $c$ could be determined in precision experiments with pions. What is quoted here is a rough indication.}
\be
f_\pi^*/f_\pi &\approx&1/(1+c n/n_0),\ \  c \approx 0.2 \ \ {\rm for }\ \ n< n_{1/2}\,, \nonumber\\
 &\approx& 0.8 \ \ {\rm for}\ \ n\geq n_{1/2}.
 \ee
 What is noteworthy here is that while the pion decay constant falls with density according to what is expected in chiral perturbation theory up to $n_{1/2}$ -- and consistent with experiments up to $n_0$, it stops dropping at the skyrmion-half-skyrmion changeover density and stays constant up to possible chiral restoration point $n_c$\footnote{It is found on the ground of RG properties of HLS theory in medium that a fixed point called ``dialton limit fixed point" (DLFP) intervenes before reaching the chiral transition point~\cite{paeng}. Thus the theory must break down before reaching $n_c$. This should be understood when we say ``toward chiral restoration."}.
 \item The in-medium nucleon mass (or rather the in-medium $B=1$ soliton mass) drops proportionally to the pion decay constant as predicted in the large $N_c$ limit. Hence it has the behavior
 \be
m_N^*/m_N &\approx&1/(1+c n/n_0),\ \  c \approx 0.2 \ \ {\rm for }\ \ n< n_{1/2}\,, \nonumber\\
 &\approx& 0.8 \ \ {\rm for}\ \  n\geq n_{1/2}.
 \ee
 Again the remarkable feature in this prediction is that the nucleon mass stops dropping at $n_{1/2}$. This means that the nucleon mass has a major portion $\sim 0.8m_N$ that does not disappear as the quark condensate tends toward zero.
\end{enumerate}
In order to extract the properties of the vector mesons, it turns out that the skyrmion crystal as worked out in \cite{park-vento,maetalcrystal} is not sufficient.  One expects on general ground the vector meson masses to scale  as
\be
m_\rho^*/m_\rho &\approx& (f_\pi^*/f_\pi)\ \ {\rm for}\ \ n< n_{1/2}\,,\nonumber\\
&\approx& (f_\pi^*/f_\pi) (g_\rho^*/g_\rho)\ \  {\rm for }\ \ n\geq n_{1/2},
\ee
and
\be
m_\omega^*/m_\omega&\approx& (f_\pi^*/f_\pi)\ \ {\rm for}\ \ n< n_{1/2}\,,\nonumber\\
&\approx& (f_\pi^*/f_\pi) (g_\omega^*/g_\omega)\ \  {\rm for }\ \ n\geq n_{1/2}.
\ee
Here $g_\rho$ is the $SU(2)$ hidden gauge coupling for $\rho$ and  $g_\omega$ is defined similarly for $U(1)$ gauge field for the $\omega$ meson.  $U(2)$ symmetry is not assumed for $\rho$ and $\omega$. The skyrmion crystal calculation does not provide information on the couplings $g_{\rho,\omega}$. In (hidden) gauge theory, they are given by loop corrections which involve $1/N_c$ corrections. Such loop effects are evidently not captured in the crystal calculation. In fact it is not clear how to incorporate them within the skyrmion crystal calculation. For this we have to resort to other approaches available within the HLS framework. This we do as follows.

From the work of Harada and Yamawaki using the RG flow of HLS theory~\cite{HY:PR}, we have, as density  approaches the chiral restoration density $n_c$, the vector manifestation (VM for short) fixed point
\be
m_\rho^*/m_\rho\sim g_{\rho}^*/g_{\rho} \sim \la\bar{q}q\ra^*/\la\bar{q}q\ra \rightarrow 0\ \ {\rm as} \\ n\rightarrow n_c.
\ee
There is however no information on the $U(1)$ coupling $g_\omega$ from the RG analysis~\cite{HY:PR}, so we do not know the behavior of the $\omega$ meson for $n\gsim n_{1/2}$.

Consider now how mesons couple to the nucleon. Here hidden local symmetry brings in something hitherto unexpected. The effective coupling of the vector meson $V=(\rho,\omega)$ has the form
\be
g^*_{VNN}=g^*_V F^*_V.
\ee
The coupling to the nucleon picks up a density dependent function $F_V$. This function is not transparent in the crystal calculation but it seems to be implicit in it as will be shown below in connection with the symmetry energy. This function runs with density  as one can see in the one-loop RG analysis in baryon-implemented HLS~\cite{paeng}. As density approaches the ``dilaton-limit fixed point" (DLFP) $n_{dl}$, $F_\rho$ tends to zero:
\be
F_\rho^*\rightarrow 0 \ \ {\rm as}\ \  n\rightarrow n_{dl} < n_c
\ee
whereas $F_\omega^*$ does not run at one-loop order. It could possibly run at higher loops, but must scale slowly. As a first approximation, we will take it to be unscaling.

In short, as density goes above $n_{1/2}$, the $\rho$-NN coupling tends to zero because of the dropping $F_\rho$, accentuated in the vicinity of $n_c$ by the gauge coupling $g_\rho$ approaching the VM fixed point. On the other hand, the $\omega$-NN coupling may stay un-dropping until very near $n_c$ although it is not known how  the gauge coupling $g_\omega$ scales. In fact the phenomenological analysis in \cite{dongetal} requires that the $\omega$-NN coupling scale little, if at all.
\subsection{New ``BLPR" scaling}
We now incorporate the above scaling parameters into the HLS Lagrangian that can be used to compute  the properties of dense matter. This amounts to transferring the topology change to changes in the parameters of the HLS Lagrangian. This brings major modifications at high density to the scaling Lagrangian of 1991~\cite{BR91} referred to as ``old BR". The modifications come only for density $n\geq n_{1/2}$, with the properties below $n_{1/2}$ being essentially the same as in \cite{BR91}. Since the new BR we have here is gotten from recent developments that involve also Byung-Yoon Park (skyrmion crystal) and Hyun Kyu  Lee (dilaton), following \cite{LR:flavor-symmetry}, we refer to this new BR as ``BLPR scaling." We will argue that the translation of the topology change into parameter change of the Lagrangian has a support in the structure of the symmetry energy.

%The BLBR scaling reads ...
\section{Effects on Tensor Forces}
We shall now apply the above scaling properties to the symmetry energy of asymmetric nuclear matter and show that the translation is at least qualitatively validated.

Since the in-medium nucleon remains heavy within the density range involved, we can use the NR approximation for the nucleon coupling to the pion and the $\rho$ and write down the tensor forces implied by the HLS Lagrangian. They take the form
\begin{eqnarray}
V_M^T(r)&&= S_M\frac{f_{NM}^2}{4\pi}m_M \tau_1 \cdot \tau_2 S_{12}\nonumber\\
&& \left(
 \left[ \frac{1}{(m_M r)^3} + \frac{1}{(m_M r)^2}
+ \frac{1}{3 m_Mr} \right] e^{-m_M r}\right),
\label{tenforce}
\end{eqnarray}
where $M=\pi, \rho$, $S_{\rho(\pi)}=+1(-1)$. Note that the tensor forces come with an opposite sign between the pion and $\rho$ tensors.

It is found to be a good approximation to have the pion tensor unaffected by the density for the range of density we are concerned with. As for the $\rho$ tensor, apart from the scaling mass $m_\rho^*$, it is the scaling of $f_{N\rho}$ that is crucial. Written in terms of the parameters of the baryon HLS Lagrangian (BHLS), we have the ratio
\be
R\equiv \frac{f_{N\rho}^*}{f_{N\rho}}\approx \frac{g_{\rho NN}^*}{g_{\rho NN}}\frac{m_\rho^*}{m_\rho}\frac{m_N}{m_N^*}\ .
\ee
It follows from the old and BLPR scaling relations that
\be
R&\approx& 1\ \ \ {\rm for}\ \ \ 0\lsim n \lsim n_{1/2}\label{R1} \\
& \approx& \left(\frac{F_\rho^*}{F_\rho}\right){\Phi}^2 \ \ {\rm for} \ \  n_{1/2}\lsim n \lsim n_c\label{R2}
\ee
where we have defined a scaling function that has -- in the chiral limit -- the VM fixed point,
\be
\Phi\approx g^*_\rho/g_\rho\rightarrow 0 \ \ {\rm as} \ \ n\rightarrow n_c.
\ee
We note that the ratio $R$ will be strongly suppressed for $n>n_{1/2}$: In addition to the suppression by the factor $\Phi^2$ that drops in going to the vector manifestation fixed point, the approach to the dilaton-limit fixed point $F^*_\rho=0$ (see Eq.~(\ref{fixed})) would make  $R$ drop {\it faster} approaching the VM fixed point. This would make the $\rho$ tensor killed extremely rapidly in the region $n>n_{1/2}$.

\begin{figure}[ht!]
\begin{center}
\includegraphics[height=5.3cm]{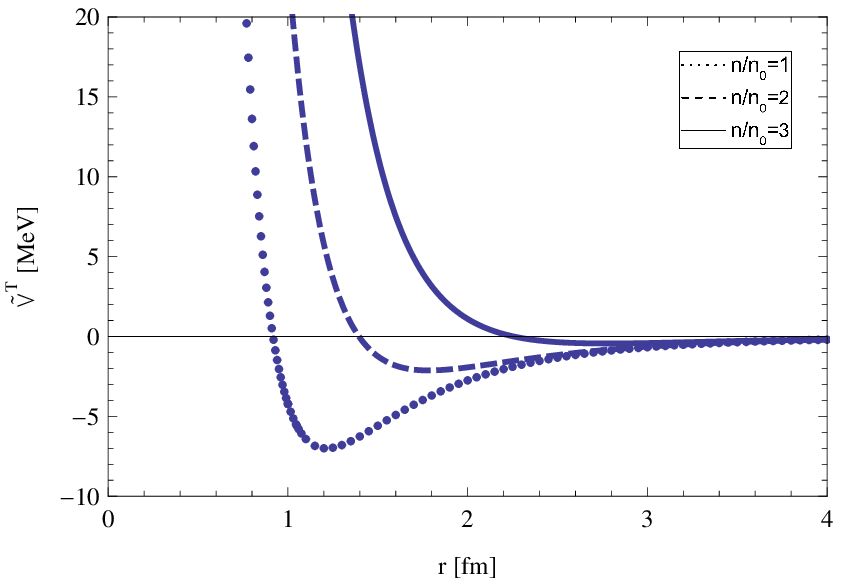}
\includegraphics[height=5.3cm]{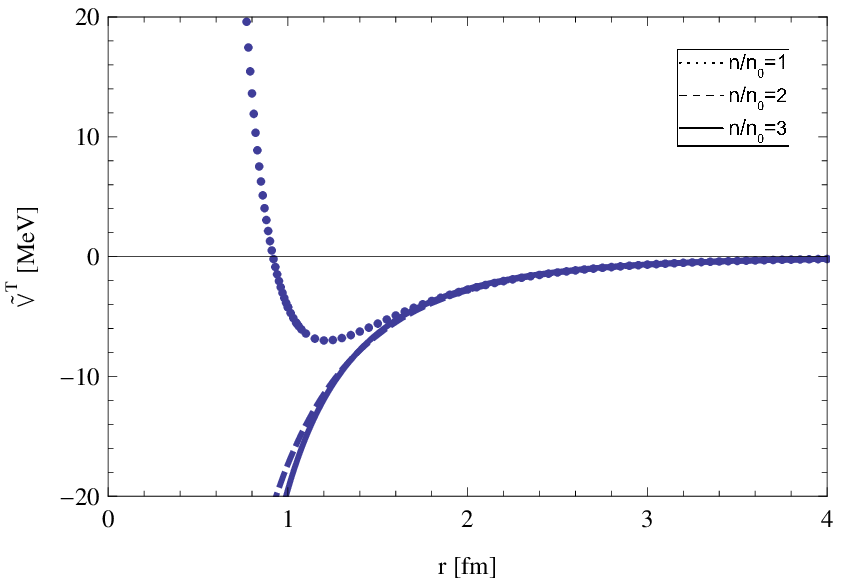}
%\vskip 0.5cm
\caption{Sum of $\pi$ and $\rho$ tensor forces 
%$\tilde{V}^T\equiv (\tau_1 \cdot \tau_2 S_{12})^{-1} (V_\pi^T +V_\rho^T)$ 
in units of MeV for densities $n/n_0$ =1 (dotted), 2 (dashed) and 3 (solid) with the ``old scaling" $\Phi \approx  1-0.15 n/n_0$ and $R\approx 1$ for all $n$ (left panel) and with the ``new scaling," $\Phi_I \approx 1-0.15 n/n_0$  with  $R\approx 1$ for $n<n_{1/2}$ and $\Phi_{II}\approx \Phi_I$ and $R\approx \Phi_{II}^2$ for $n>n_{1/2}$, assuming $ n_0<n_{1/2}<2n_0$ (right panel). For simplicity we set $F_\rho^*/F_\rho=1$.}\label{tensor}
\end{center}
\end{figure}

The drastic change in the tensor forces going over to the half-skyrmion phase is illustrated in Fig.~\ref{tensor}.  Here scaling parameters that are considered to be reasonable are used. The general feature will not be modified for small variation of parameters in the vicinity of what's picked.
It is seen that with the old BR where the skyrmion-half-skyrmion phase change is not taken into account, the net tensor force decreases continuously in density with the attraction vanishing at $n\lsim 3n_0$ for the given parameters. In a stark contrast, with the BLPR, the $\rho$ tensor force nearly disappears at $n\sim 3n_0$ leaving only the pion tensor active, making the strength of the tensor force increase at $n_{1/2}$. Taking into account the factor $F_\rho^*/F_\rho$ with its fixed point would make the suppression of the $\rho$ tensor even more dramatic than what's given in Fig.~\ref{tensor}.
\section{Breakdown of Mean Field Theory in the Half-Skyrmion Phase: A Conjecture}
One of the most direct and immediate consequences of the change in tensor forces is on the symmetry energy $S$,  the coefficient multiplying the factor $\alpha^2=[(N-P)/(N+P)]^2$ (where $N(P)$ is the number of neutrons(protons)) in the expansion of the energy per particle $E$ of asymmetric nuclear system. Assuming with others~\cite{BM,xu-li,polls} that near the equilibrium nuclear matter density, the symmetry energy is dominated by the tensor forces, we have, using closure approximation~\cite{BM},
\be
S\sim \frac{12}{\bar{E}}\la V_T^2(r)\ra\label{BM}
\ee
where $\bar{E}\approx 200$ MeV is the average energy typical of the tensor force excitation and $V_T$ is the radial part -- with the scaling factor $R$ taken into account -- of the net tensor force. One can see from Fig.~\ref{tensor} that $S$ decreases until $n_{1/2}$ but then turns over and increases as the pion tensor takes over. This predicts a cusp at $n_{1/2}$. This cusp structure is precisely reproduced in the skyrmion description as shown in \cite{LPR-halfskyrmion}. In the skyrmion crystal, the symmetry energy arises as  a $1/N_C$ term coming from the collective quantization. It is subleading in $N_c$, but it is well-defined and expected to be robust, independent of dynamical details.

We now argue that the above cusp structure is lost if we apply the mean-field approximation to the Lagrangian BHLS we are using.  In order to do the mean field which is justified  near $n_0$ as mentioned, we need to implement a scalar field to the baryon HLS Lagrangian.  Otherwise we cannot have nuclear saturation. With the dilaton scalar field $\chi$ introduced via trace anomaly as pseudo-Goldstone field to encode spontaneously broken conformal symmetry,  one can do the mean field calculation to describe matter at the equilibrium density $n_0$. With a suitable scaling in the Lagrangian parameters that satisfy BR scaling, the model works fairly well at near $n_0$~\cite{songetal}.  Here the relevant scalar is a chiral singlet or  in a more realistic description, dominantly chiral singlet with small non-singlet admixtures. However as density increases toward the dilaton-limit fixed point, in order to exhibit correct chiral symmetry, the same baryonic HLS should flow to a Gell-Mann-L\'evy-type Lagrangian in which the scalar field becomes the fourth component of the chiral four vector of $SU(2)_L\times SU(2)_R$. This means that at some density, say, above $n_{1/2}$, the dilaton limit fixed point is approached~\cite{paeng}:
\be
F^*_\rho\rightarrow 0, \ \ n\rightarrow n_{dl} < n_c.\label{fixed}
\ee
Doing mean field with this Lagrangian gives
\be
S\sim (F^*_\rho/F_\pi^*)^2 n.
\ee
Since $F^*_\pi$ stays more or less constant up to the would-be critical density $n_c$, the scaling of the symmetry will be dictated by $F_\rho^*$. As $F_\rho^*$ drops to zero approaching the DLFP $n_{dl}$, $S/n$ will continue to decrease after $n_{1/2}$. This is in disagreement with both the skyrmion prediction~\cite{LPR-halfskyrmion} and the BLPR-scaling-implemented effective field theory prediction (\ref{BM})~\cite{paeng}, both of which in some ways, seem to capture many-body correlation effects that go beyond the mean field. Given that the dilaton-implemented BHLS Lagrangian is to {\em describe simultaneously} both $n\sim n_0$ and $n\sim n_{dl}$, it must be that the mean field approximation breaks down as one enters into the half-skyrmion phase. This leads to our conjecture:The skyrmion-half-skyrmion transition that involves a topology change is tantamount to a Fermi-liquid-non-Fermi liquid transition. There are similar cases in condensed matter physics where the change from Fermi-liquid to non-Fermi liquid can be precisely formulated in terms of thermodynamic properties. For instance in  \cite{condensed matter}, in  a high pressure study of certain metallic state in the presence of topologically distinct spin textures (i.e., hedgehogs) it is seen that spin correlations with non-trivial topology can drive a breakdown of Fermi liquid structure. In our case, although not evident, the topology associated with skyrmions describes, near the equilibrium nuclear matter, that Fermi-liquid structure is destroyed  going into the half-skyrmion phase.  The deviation from canonical Fermi-liquid structure manifests itself in that the quasiparticle structure of dominant single-particle propagator is destroyed.  As will be elaborated below, this non-quasiparticle structure is seen in the EoS for dense baryonic matter in a formalism in which higher-order correlations are taken into account~\cite{dongetal}.
\section{Prediction for the EoS of Compact-Star Matter}
In order to confront nature, we need to perform quantum calculations, going beyond the mean field. The features discussed above are quasi-classical and qualitative, so what we have obtained have to be incorporated into a more quantitative formalism.

Our procedure is to inject the parameters scaling according to the BLPR into the baryon hidden local symmetric Lagrangian and since the quasiparticle approximation is no longer reliable, to do higher-order many-body calculations for densities $n>n_{1/2}$. How to go about doing this is not precisely formulated up to date. In \cite{dongetal}, this strategy is implemented in the $V_{lowk}$-EFT approach, by relying on available experiments for guidance.

In the density regime $n\leq n_0$,  the parameters are controlled by available data. It seems reasonable to assume -- that we do -- that this procedure holds (at least) up to $n_{1/2}$ which is not far above $n_0$. This density regime will be referred to as Region I. It is in the density regime $n>n_{1/2}$ -- that we refer to as Region II -- that the topology change brings in significant modifications in the $V_{lowk}$-implemented EFT approach.

The predictions made~\cite{dongetal} with the scalings in Region II as sketched above suitably\footnote{There is no model-independent information on the precise scaling behavior. It is largely guided with the limited source from experiments. However what has been used in \cite{dongetal} has been recently given a support from different approaches anchored on hidden local symmetry~\cite{paeng}.} taken into account in the $V_{lowk}$-EFT are summarized in the figures given below.

In Fig.~\ref{figure1} are given the EoS for symmetric nuclear matter and the symmetry energy $S$.
\begin{figure}[here]
\begin{center}
\scalebox{0.30}{\includegraphics[angle=-90]{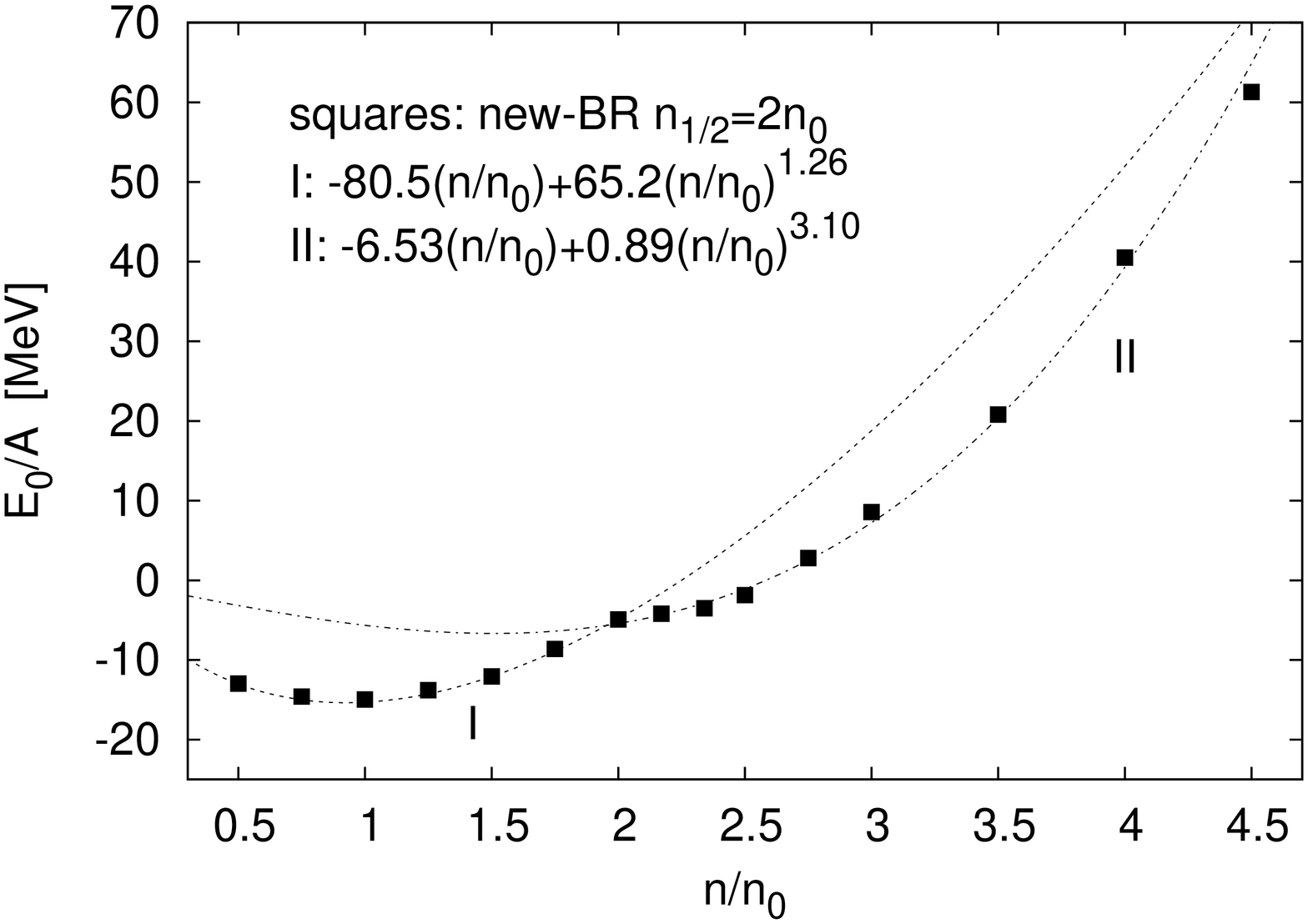}}\scalebox{0.30}{\includegraphics[angle=-90]{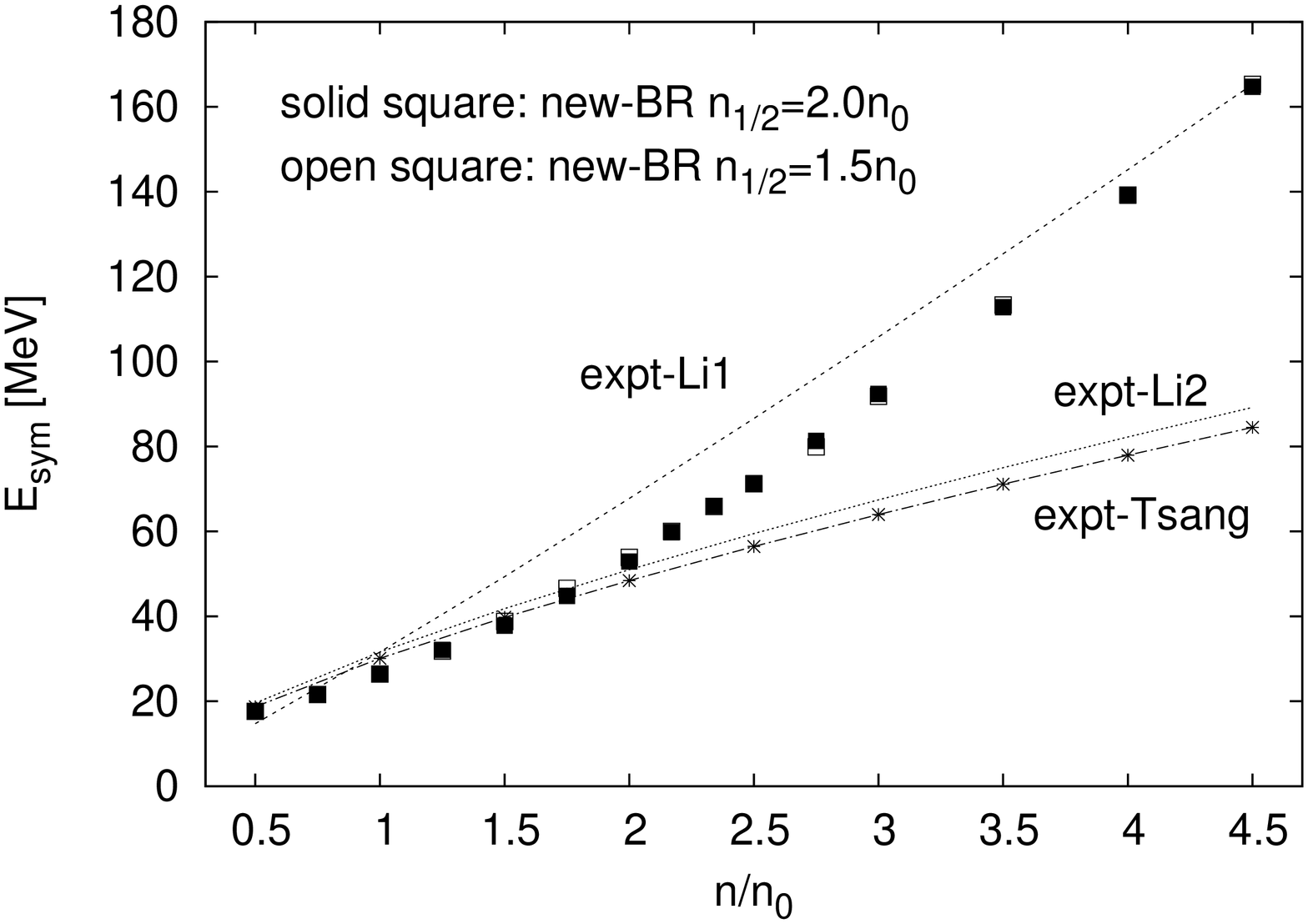}}
\end{center}
\caption{EoS for symmetric nuclear matter $E_0/A$ with polytrope fits of the $n_{1/2}=2n_0$ and symmetry energy $S$.  }\label{figure1}
\end{figure}
%\begin{figure}[here]
%\begin{center}
%\scalebox{0.34}{\includegraphics[angle=-90]{figure2.eps}}
%\end{center}
%\caption{Nuclear symmetry energy: Comparison with the empirical upper (expt-Li1)
%and lower (expt-Li2) constraints of Li et al. \cite{li05} and the empirical results of
%Tsang et al. (expt-Tsang). }\label{figure2}
%\end{figure}
It is noteworthy that the onset of half-skyrmion phase makes the EoS of symmetric nuclear matter {\it softer} whereas the symmetry energy becomes {\it stiffer}. This feature can be easily understood in terms of  the drastic change in the tensor forces associated with the phase change and how nuclear forces enter in EoS. In condensed matter, the changeover from Fermi liquid to non-Fermi liquid linked to a topology change modifies the temperature dependence in the resistivity, for instance~\cite{condensed matter}. It would be interesting in the case we are dealing with to establish more precise scaling properties in density as one goes from one phase to the other. One can already see a hint in a polytrope fit for $E_0/A$~\cite{dongetal}. In Fig.~\ref{figure1} is seen that  the polytrope fit requires two different forms with quite different density dependence changing over at $n_{1/2}$.

\begin{figure}[here]
\begin{center}
%\scalebox{0.34}{\includegraphics[angle=-90]{figure3.eps}}
\scalebox{0.30}{\includegraphics[angle=-90]{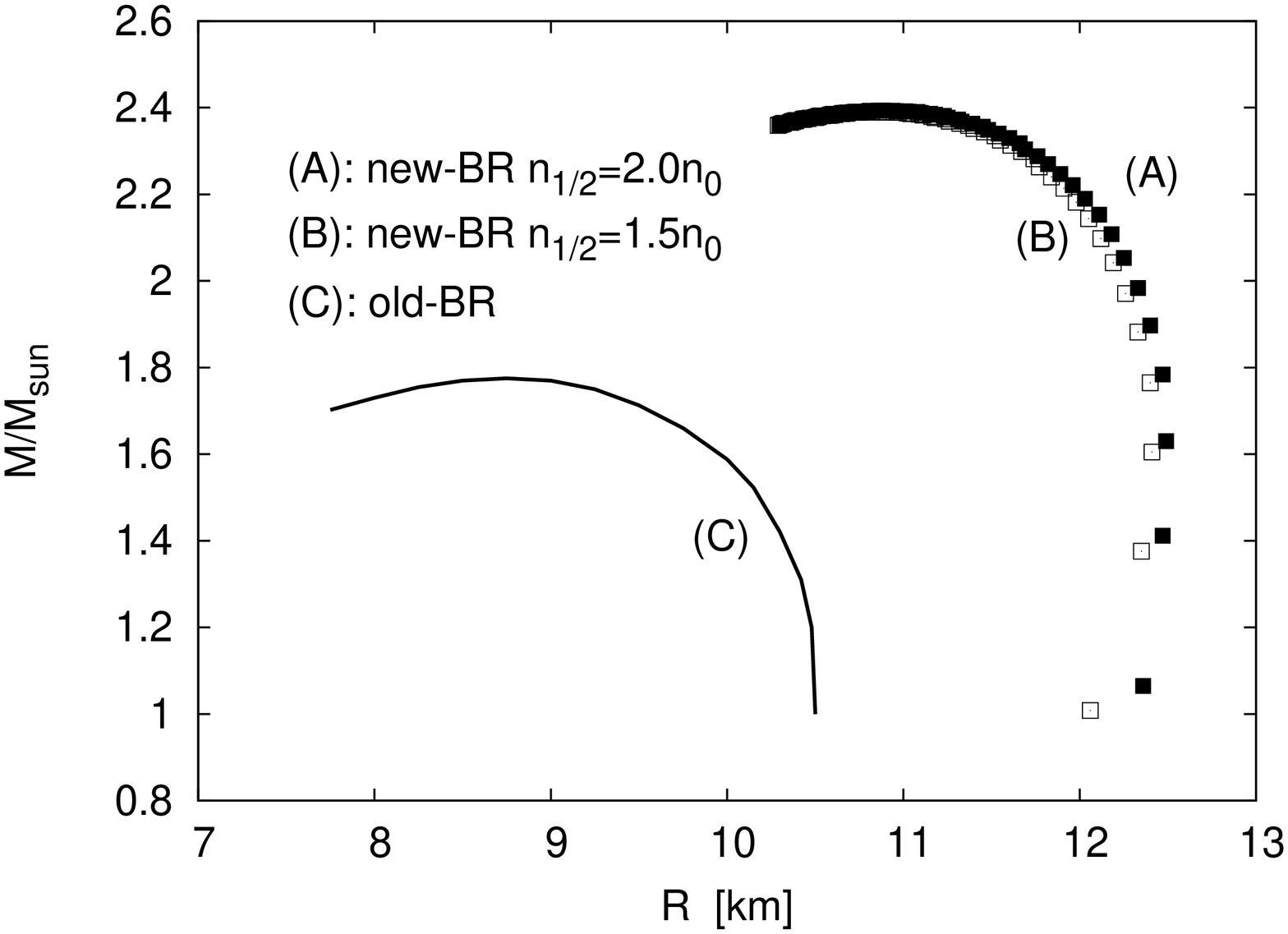}}\scalebox{0.30}{\includegraphics[angle=-90]{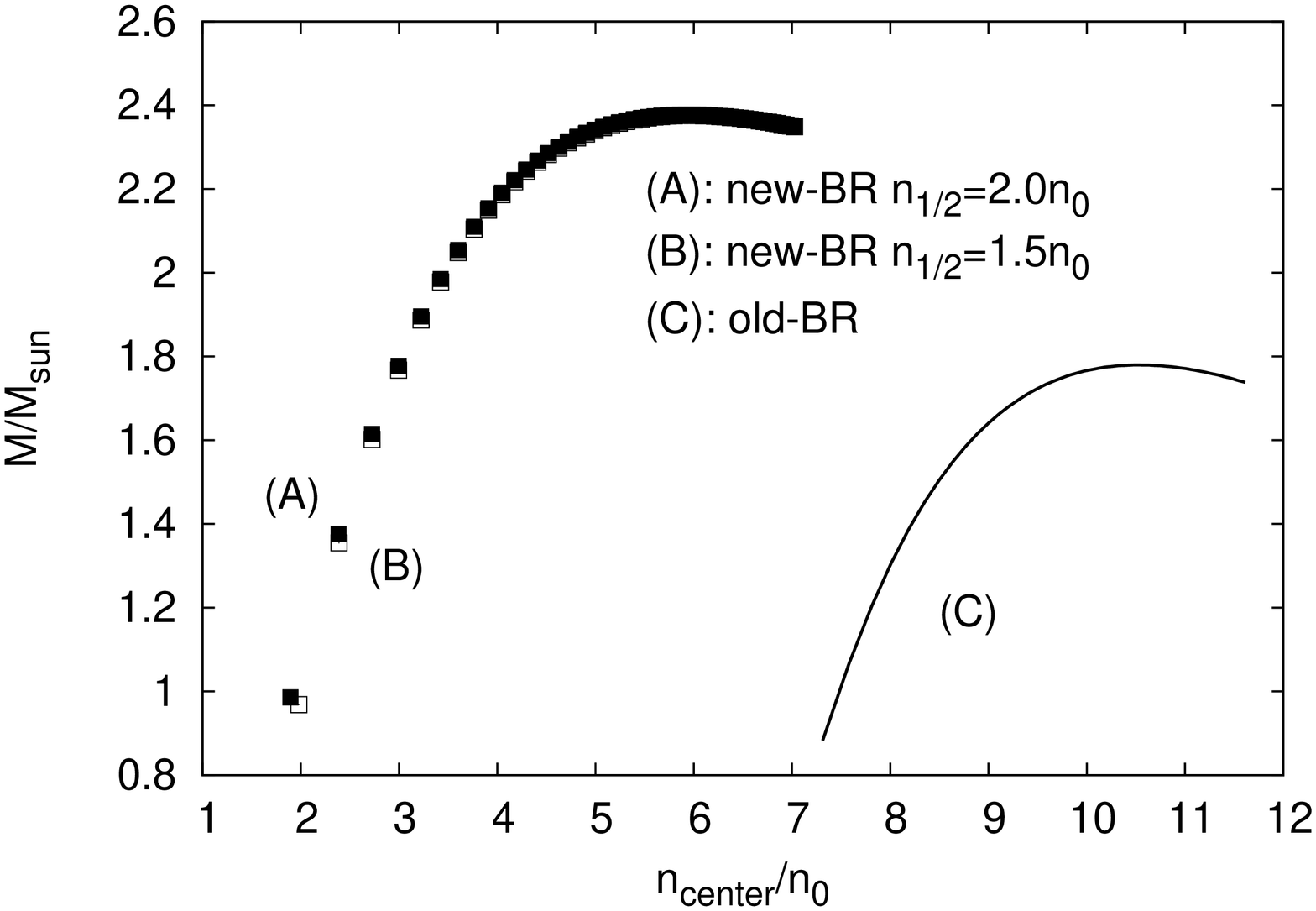}}
\end{center}
\caption{Mass-radius trajectories (left panel) and central densities (right panel) of
neutron stars calculated with topology change using $n_{1/2}$ = 2.0 (A) and 1.5$n_0$ (B) and without topology change (C).}\label{figure3}
\end{figure}

 To give an idea as to how this model can accommodate massive neutron stars, we show the results from \cite{dongetal} without going into details of the choice of scaling etc.  Fig.~\ref{figure3} depicts the mass-radius trajectories and maximum densities of neutron stars for two values of $n_{1/2}$ that give the rough range predicted by HLS skyrmions of \cite{maetalcrystal}. The results are more or less insensitive to the precise value of $n_{1/2}$.  What is striking is the effect of topology change for both the maximum mass and radius, with the massive star of $\sim 2 M_\odot$ having a bigger radius predicted with the  topology change than a less massive star of $\sim 1.6 M_\odot$ without. There is also a marked difference in the maximum density in the interior of the star. Without the topology change the EoS is soft and hence packs a bigger -- by a factor of 2 -- density.

\section{Origin of the Nucleon Mass}
\begin{figure}[ht]
\centering
\includegraphics[scale=0.4]{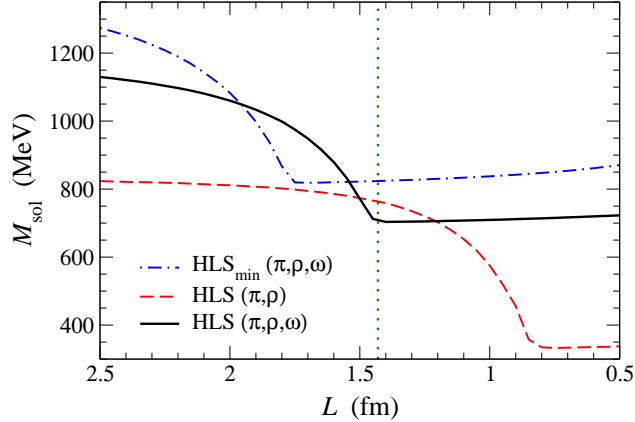}
\caption[]
{Crystal size $L$ dependence of soliton mass (large $N_c$ proton mass) with the FCC crystal background. Density increases to the right. The solid curve is the result of the model adopted in \cite{maetalcrystal}. Others represent results when different parameterizations than what's given by the master formula are used.}
\label{Nmass}
\end{figure}

In confronting the EoS compatible with the observed properties of the $\sim 2$ solar neutron star, it was crucial in \cite{dongetal} to keep the nucleon mass non-scaling up to the maximum density relevant, $\sim 5.5 n_0$. This feature is, a posteriori, supported by the skyrmion crystal calculation~\cite{maetalcrystal}  as seen in Figure \ref{Nmass}. This feature is also reproduced in an RG-analysis with baryon HLS Lagrangian~\cite{paeng}. If the nucleon mass were to remain unchanged as density approaches first the DL fixed point and then VM fixed point, this would imply that there is a large part of the nucleon mass that remains ``unmelted" when the quark condensate melts. In terms of the nucleon mass expressed schematically as $m_N=m_0+\Delta m(\la\bar{q}q\ra)$, one can state that $\Delta m\rightarrow 0$ but $m_0$ remains unaffected when $\la\bar{q}q\ra\rightarrow 0$ . This is reminiscent of the parity-doublet nucleon model with a constant $m_0$~\cite{paritydoublet} that remains when chiral symmetry is restored. In this model, $m_0$ is a chiral-invariant and could be considered as an intrinsic quantity of QCD. The constant $m_0$ found in the skyrmion crystal,  however, appears to reflect an emergent rather than intrinsic symmetry. In fact, the half-skyrmion crystal phase has a higher spatial symmetry than the skyrmion phase and it seems likely that $m_0$ arises from collective in-medium correlations.

As pointed out in \cite{dongetal}, the presence of $m_0$ in the nucleon mass brings tension in the application of the constituent quark model (CQM) to in-medium hadron masses. The CQM has a strong support by the large $N_c$ consideration of QCD~\cite{CQM-weinberg}. However while it can be applied to, say, the $\rho$ meson, with the constituent quark mass subject to the vector manifestation fixed point, it cannot be to the baryon mass with a large $m_0$. This dichotomy of the meson and baryon in-medium masses is evident when the baryon is considered as a chiral bag~\cite{BR-littlebag} and could perhaps be understood in terms of large $N_c$ QCD recently formulated by Kaplan~\cite{kaplan}.
\section{Experimental Tests}
Since the effect on the tensor forces is the most striking, the measurement of the symmetry energy will be one of the primary targets for experimental tests of the theory developed in this note. The BLPR-implemented-EFT approach~\cite{dongetal} backed by the skyrmion crystal calculation~\cite{maetalcrystal} and the RG analysis~\cite{paeng} indicates that a slope change at $n_{1/2}$ in the symmetry energy could be measured in forthcoming experiments at RIB-type accelerators as well as at FAIR/GSI and other accelerators. Another direction is to zero-in on the structure of tensor forces in neutron-rich nuclei. In an elegant analysis, Otsuka and collaborators showed that the ``bare" tensor forces are left un-renormalized in certain channels by both short-range correlations and core polarizations~\cite{otsuka-NR}. It is found in particular in the monopole matrix elements that the intrinsic tensor forces remain unscathed by nuclear correlations whereas other components can be massively renormalized. This strongly suggests that the bare tensor force strength can be pinned down in nuclear medium by looking at processes sensitive to the monopole matrix element such as single-particle shell evolution~\cite{shell evolution} and if there is any medium effect reflecting the {\it fundamental} change in the chiral condensate due to density-induced vacuum change, it should show up in a pristine way. If $n_{1/2}$ is close to $n_0$, RIB-type accelerators could provide a hint to this phase change associated with a topology change.
\subsection*{Acknowledgments}
We are grateful for valuable discussions with Tom Kuo,  Won-Gi Paeng,  Byung-Yoon Park and Chihiro Sasaki with whom important part of the research discussed here was performed. This work was partially supported by the WCU project of the Korean Ministry of Educational Science and Technology (R33-2008-000-10087-0).

\end{document}